\documentclass{article}
\usepackage{natbib}
\usepackage{graphicx}
\def\A{{\tt A}}
\def\C{{\tt C}}
\def\G{{\tt G}}
\def\T{{\tt T}}
\def\ApA{{\tt ApA}}
\def\ApC{{\tt ApC}}

\def\CpA{{\tt CpA}}
\def\CpC{{\tt CpC}}
\def\CpG{{\tt CpG}}

\def\GpA{{\tt GpA}}
\def\GpC{{\tt GpC}}
\def\GpG{{\tt GpG}}
\def\GpT{{\tt GpT}}
\def\TpA{{\tt TpA}}
\def\TpC{{\tt TpC}}
\def\TpG{{\tt TpG}}
\def\TpT{{\tt TpT}}
\def\ra{\rightarrow}
\def\statGC{stationary GC-content}
\def\rat{$\ra$}

\def\tabi{\begin{table}[bth]
\begin{center}
\footnotesize
\begin{tabular}{r|c|c|c|c}
&6 parameter
&7 parameter
&8 parameter
&9 parameter
\\
&model
&model
&model
&model
\\
\hline
\A:\T\rat\C:\G
&0.012
&0.012
&0.011
&0.007
\\
\A:\T\rat\T:\A
&0.010
&0.011
&0.011
&0.011
\\
\C:\G\rat\G:\C
&0.016
&0.016
&0.012
&0.012
\\
\C:\G\rat\A:\T
&0.015
&0.014
&0.014
&0.014
\\
\A:\T\rat\G:\C
&0.036
&0.036
&0.036
&0.036
\\
\C:\G\rat\T:\A
&0.158
&0.059
&0.060
&0.060
\\
\hline
\CpG\rat\CpA/\TpG
&
&0.618
&0.627
&0.624
\\
\CpG\rat\CpC/\GpG
&
&
&0.029
&0.029
\\
\TpT/\ApA\rat\TpG/\CpA
&
&
&
&0.013
\\
\hline
\statGC 
&0.213
&0.341
&0.340
&0.339
\\
\hline
$-2 \log\lambda$ 
&
&7.7$\cdot 10^6$
&1.3$\cdot 10^5$
&9.6$\cdot 10^4$
\end{tabular}
\caption{
\footnotesize
\label{tab1}Estimates for substitution frequencies for nested models
of nucleotide substitution in human AluSx repeats. Given are the substitution
frequencies per bp in the time span after the insertion of the AluSx repeats
into the human genome. In the last row we note the $-2\log\lambda$ where $\lambda$ is the
likelihood ratio of the model and the one with one less parameter in the column
to the left.}
\end{center}
\end{table}
}

\def\tabii{\begin{table}[bht]
\begin{center}
\footnotesize
\begin{tabular}{r|c|c|c|c}
&6 parameter
&7 parameter
&8 parameter
&9 parameter
\\
&model
&model
&model
&model
\\
\hline
\A:\T\rat\C:\G
&0.024
&0.025
&0.026
&0.026
\\
\A:\T\rat\T:\A
&0.041
&0.041
&0.041
&0.041
\\
\C:\G\rat\G:\C
&0.037
&0.036
&0.036
&0.023
\\
\C:\G\rat\A:\T
&0.029
&0.029
&0.028
&0.028
\\
\A:\T\rat\G:\C
&0.073
&0.074
&0.046
&0.046
\\
\C:\G\rat\T:\A
&0.151
&0.111
&0.105
&0.107
\\
\hline
\CpG\rat\CpA/\TpG
&
&0.274
&0.331
&0.328
\\
\CpA/\TpG\rat\CpG
&
&
&0.100
&0.097
\\
\CpG\rat\CpC/\GpG
&
&
&
&0.096
\\
\hline
\statGC 
&0.349
&0.374
&0.335
&0.337
\\
\hline
$-2 \log\lambda$ 
&
&2.9$\cdot 10^5$
&1.6$\cdot 10^5$
&1.1$\cdot 10^5$
\end{tabular}
\caption{
\footnotesize
\label{tab2}Estimates for substitution frequencies for nested models
of nucleotide substitution in DANA repeats from {\em Danio rerio}.}
\end{center}
\end{table}
}

\def\tabiii{\begin{table}[htb]
\begin{center}
\footnotesize
\begin{tabular}{r|c|c|c|c}
&6 parameter
&7 parameter
&8 parameter
&9 parameter
\\
&model
&model
&model
&model
\\
\hline
\A:\T\rat\C:\G
&0.038
&0.038
&0.038
&0.038
\\
\A:\T\rat\T:\A
&0.052
&0.045
&0.045
&0.045
\\
\C:\G\rat\G:\C
&0.034
&0.034
&0.034
&0.034
\\
\C:\G\rat\A:\T
&0.074
&0.074
&0.074
&0.074
\\
\A:\T\rat\G:\C
&0.052
&0.052
&0.052
&0.047
\\
\C:\G\rat\T:\A
&0.108
&0.108
&0.098
&0.098
\\
\hline
\TpA\rat\TpT/\ApA
&
&0.029
&0.028
&0.028
\\
\TpC/\GpA\rat\TpT/\ApA
&
&
&0.036
&0.035
\\
\GpT/\ApC\rat\GpC
&
&
&
&0.021
\\
\hline
\statGC 
&0.330
&0.330
&0.328
&0.326
\\
\hline
$-2 \log\lambda$ 
&
&853
&592
&40
\end{tabular}
\caption{
\footnotesize
\label{tab3}Estimates for substitution frequencies for nested models of nucleotide
substitution in DNAREP1\_DM transposable element from {\em Drosophila melanogaster}.}
\end{center}
\end{table}
}

\author{Peter F. Arndt${}^{1*}$ and Terence Hwa${}^2$\\[5mm]
${}^1$ Max Planck Institute for Molecular Genetics, \\
Ihnestr. 73, 14195 Berlin, Germany\\[1mm]
${}^2$ Center for Theoretical Biological Physics, 
\\
UC San Diego, 
9500 Gilman Drive, La Jolla, CA 92093-0374
\\[3mm]
${}^*$ To whom correspondence should be addressed.
}

\title{Identification and Measurement of Neighbor Dependent Nucleotide Substitution Processes}
\begin{document}
\maketitle

\begin{abstract}
\mbox{}\\\noindent
{\bf Motivation:} 
The presence of neighbor dependencies generated a specific pattern of
dinucleotide frequencies in all organisms.  Especially, the
CpG-methylation-deamination process is the predominant substitution process in
vertebrates and needs to be incorporated into a more realistic model for
nucleotide substitutions. 
\\\noindent
{\bf Results:} 
Based on a general framework of nucleotide substitutions we develop a method
that is able to identify the most relevant neighbor dependent substitution
processes, measure their strength, and judge their importance to be included
into the modeling. Starting from a model for neighbor independent nucleotide
substitution we successively add neighbor dependent substitution processes in
the order of their ability to increase the likelihood of the model describing
given data. The analysis of neighbor dependent nucleotide substitutions in
human, zebrafish and fruit fly is presented.
\\\noindent
{\bf Availability:} A web server to perform the presented analysis is
publicly available at:
http://evogen.molgen.mpg.de/server/substitution-analysis . 
\\\noindent
{\bf Contact:} arndt@molgen.mpg.de 

\end{abstract}

\section{Introduction}
The identity of the neighboring nucleotide can have a drastic influence on the
mutation rates of a nucleotide. A well-known and studied example of this fact
is the increased mutation of cytosine to thymine in \CpG\ dinucleotides in
vertebrates \citep{Co78, RR80}. This process is triggered by the methylation of
cytosine in \CpG\, followed by deamination, and mutation from \CpG\ to \TpG\ or
\CpA\ (on the reverse strand). Due to this process the number of \CpG\ is
decreased while the number of \TpG\ and \CpA\ is larger than expected from
independently evolving nucleotides. Most of the deviant dinucleotide odds
ratios (dinucleotide frequencies normalized for the base composition) in the
human genome can be explained by the presence of the \CpG\ methylation
deamination process \citep{ABH02}. Biochemical studies in the 1970s already
compared these odds ratios for different genomes and different fractions of
genomic DNA \citep{Ru76, RS77} and concluded that these ratios are a remarkably
stable property of genomes. In the following Karlin and coworkers \citep{CB95,
KM97, KMC97} elaborated and expanded these observations, showing that the
pattern of dinucleotide abundance constitutes a genomic signature in the sense
that it stable across different parts of a genome and generally similar between
related organisms. Since this signature is also present in non-coding and
intergenic DNA it is very promising to study neighbor dependent mutation and
fixation processes (we refer to the effective process as the substitution
process) to understand the evolution of neutral DNA. 
However, to pursue on this track new models for nucleotide
substitutions that extends those which only capture neighbor independent
nucleotide substitutions (see  \citep{LioGoldman} for a review) have to be
formulated (see also \citep{ABH02, Haussler, LH04}). 

Recently a framework to include such neighbor dependent processes has been
introduced \citep{ABH02}.  The framework itself is capable to include any type
of neighbor dependent process and was already successfully applied to model the
\CpG\ methylation deamination process in vertebrates \citep{APH03}.  Although
these models are mathematically more complicated they however allow a
quantitative analysis of neighbor dependent processes and to make reliable
estimations on other properties e.g. the stationary GC-content.  Here we will
extend this framework and discuss the inclusion of more neighbor dependent
substitutions and how one can infer their relevance without prior knowledge on
the underlying biochemical processes.  In vertebrates the \CpG\ methylation
deamination process is the predominant nucleotide substitution process. Its
rate is about 40 times higher than this of a transversion and its history can
actually reconstructed for the last 250 Myr \citep{APH03}. One reason for this
substitution frequency being so high is that in vertebrates \CpG\ methylation is also
used in gene regulation, 
as methylated regions of the genome are not transcribed.
Consequently, \CpG's in these regions often mutate. We
know already that also other vertebrates use methylation in the same way but do
not know about the quantitative extent their genomes are methylated. The
situation is still rather unclear in other kingdoms of life. Although we
clearly see signatures of neighbor dependent substitution processes, we do not
know the responsible processes and their rates.

To present our method we study neighbor dependent substitutions in human ({\em
Homo sapiens}), zebrafish ({\em Danio rerio}) and fruit fly ({\em Drosophila
melanogaster}). In all these studies we first try to model the observed
nucleotide substitutions with a model which does not include any neighbor
dependent nucleotide substitutions (12 free rate parameters) and then ask the
question which neighbor dependent substitution process one would have to
include to describe the observed data best. The idea is to capture the most of
the observed substitutions by single nucleotide substitutions independent of
the neighboring bases and then to include neighbor dependent substitutions one
by one to generate a better model with the least number of parameters.
Processes are added in the order of their ability to describe the observed data
better. Naturally, the addition of any further process (together with one rate
parameter) into a model will increase the likelihood of this model to describe
the observed data. In order not to over-fit the data we use a likelihood ratio
test to judge whether the addition of further process is justified.  The
strength of our approach is to come up with a model with fewer parameters that
still captures the essential neighbor dependent nucleotide substitution
processes. This prevents over-fitting the model to given data and eases the
quantitative estimation of a smaller number of parameters.

The rest of the paper organizes as follows. In the next section we will
describe details of our method. There is no need to implement the described
procedure for readers who want to analyze their own sequences, since we are
running a public web server at
{http://evogen.molgen.mpg.de/server/substitution-analysis}.  At this site one is
able to upload sequence data and perform the presented analysis. First
applications of such an analysis will be presented in the results section.

\section{Method}

\subsection{The substitution model}

In total there are 12 distinct neighbor independent substitution processes of 
a single nucleotides by another; four of them are so-called transitions that
interchange a purine with a purine or a pyrimidine with a pyrimidine. The
remaining eight processes are the so-called transversions that interchange a
purine with a pyrimidine and vice versa. The rates of these processes, $\alpha\ra\beta$, will be
denoted $r_{\alpha\beta}$, where $\alpha,\beta\in\{\A,\C,\G,\T\}$ denote a
nucleotide. On top of these 12 processes we want to consider also neighbor
dependent processes of the kind $\kappa\lambda\ra\kappa\sigma$ and
$\kappa\lambda\ra\sigma\lambda$
where the right or left base of a
di-nucleotide changes, respectively. There might be several of those processes
present in our model, their rates will be denoted by $r_{\kappa\lambda\kappa\sigma}$ or
$r_{\kappa\lambda\sigma\lambda}$ . We do not consider
processes where both nucleotides of a dinucleotide change at the same time. In
vertebrates, the most important neighbor dependent process to consider is the
substitution of cytosine in \CpG\ resulting in \TpG\ or \CpA. Its rate is
about 40 times higher than this of a transversion \citep{APH03}. This process is
triggered by the methylation and subsequent deamination of cytosine in \CpG\
pairs. It is commonly (and erroneously) assumed that this process only affects
\CpG\ dinucleotides. However, this is not the case as it has been shown
\citep{ABH02}.

The model is parameterized by the substitution rates and the length of
the time span,~$dt$, the respective substitution processes acted upon the sequence,
which would in our case be the time between the observation of an ancestral
sequence and its daughter sequence,~$T$. We have the freedom to rescale time and
measure it in units of $T$. In this case, the time span is $dt=1$ and with 
this choice  the substitution
rates are equal to the substitution frequencies giving the number of nucleotide
substitutions per bp. In the simplest case our model includes
neighbor independent processes only and is parameterized by 12 substitution
frequencies. For each additional neighbor dependent process we gain one
additional parameter. The set of all these substitution frequencies will be
denoted by $\{r\}$. The number of parameters can actually be reduced by a factor of two
when one considers substitutions along neutrally evolving DNA. In this case we
cannot distinguish the two strands of the DNA and therefore the substitution
rates are reverse complement symmetric, e.g. the rate for the substitution \C\rat\A\ is
equal to the rate for the substitution \G\rat\T\ (in the following we will denote this
process by $\C:\G\ra\A:\T$, for the rates we have $r_{\C\A}=r_{\G\T}$).

In order to facilitate the subsequent maximum likelihood analysis we need to
compute the probability,~$P_{\{r\}}(\cdot\beta\cdot|\alpha_1\alpha_2\alpha_3)$,
that the base $\alpha_2$ flanked by $\alpha_1$ to the left and by $\alpha_3$ to
the right, changes into the base $\beta$ for given substitution frequencies
$\{r\}$. This probability can easily calculated by numerically solving the time
evolution of the probability to find three bases $p(\alpha\beta\gamma;t)$ at
time $t$, which is given by the Master equation and can be written as the
following set of differential equations:
\begin{eqnarray}
\frac\partial{\partial t}p(\alpha\beta\gamma;t)
&=&
\sum_{\epsilon\in\{\A,\C,\G,\T\}}
\left[
r_{\epsilon\alpha}\;p(\epsilon\beta\gamma;t)
+r_{\epsilon\beta} \;p(\alpha\epsilon\gamma;t)
+r_{\epsilon\gamma} \;p(\alpha\beta\epsilon;t)
\right]
\nonumber\\
&&+
\sum_{\epsilon\epsilon'}
r_{\epsilon\epsilon'\alpha\beta}\;p(\epsilon\epsilon'\gamma;t)
+\sum_{\epsilon\epsilon'}
r_{\epsilon\epsilon'\beta\gamma}\;p(\alpha\epsilon\epsilon';t),
\label{dgl}
\end{eqnarray}
where the rate parameters with the equal initial and final state,
$r_{\alpha\alpha}$ and $r_{\alpha\beta\alpha\beta}$,
are defined by 
\begin{equation}
r_{\alpha\alpha}=-\sum_{\epsilon\neq\alpha}r_{\alpha\epsilon}
,\quad
r_{\alpha\beta\alpha\beta}=-\sum_{(\epsilon\epsilon')\neq(\alpha\beta)}r_{\alpha\beta\epsilon\epsilon'},
\nonumber
\end{equation}
and rates of neighbor dependent substitution processes not included into the
model are take to be zero.  The above definitions guarantee the conservation of
the total probability,
$\sum_{\alpha\beta\gamma}
\frac\partial{\partial t}p(\alpha\beta\gamma;t)=0
$,
since the total influx is balanced by an appropriate outflux of probability.
The first three terms on the r.h.s.~in Eq.~(\ref{dgl}) describe single
nucleotide substitutions on the three sites whereas the last two sums (which
are summed over all pairs of nucleotides) represent the neighbor dependent
processes at the sites $(1,2)$ and $(2,3)$, respectively. To describe the
evolution of three nucleotides $\alpha_1\alpha_2\alpha_3$, these differential
equations have to be solved for initial conditions of the form
\begin{equation}
p(\alpha\beta\gamma;t=0)
=\left\{
\begin{array}{cl}
1&\mbox{if }(\alpha\beta\gamma)=(\alpha_1\alpha_2\alpha_3)\\
0&\mbox{otherwise.}
\end{array}
\right.
\end{equation}
After numerically iterating the above differential equations using
the Runge-Kutta algorithm \citep{Pr92} we get the above transition probability as
\begin{equation}
P_{\{r\}}(\cdot\beta_2\cdot|\alpha_1\alpha_2\alpha_3)=
\sum_{\beta_1\beta_3}p(\beta_1\beta_2\beta_3;t=1)
\;.
\end{equation}
The above iteration has to be carried out 64 times for all possible combinations of
initial bases $\alpha_1\alpha_2\alpha_3$. After each iteration
4 of the transition probabilities
$P_{\{r\}}(\cdot\beta\cdot|\alpha_1\alpha_2\alpha_3)$
with $\beta=\A,\C,\G,$ or \T\
 can be computed.  Note, that the above
set of differential equations can easily extended to describe systems of length
$N>3$.  In this case one has to solve for $4^N$ functions
$p(\alpha_1\alpha_2\dots\alpha_N;t)$.

\subsection{Estimation of substitution frequencies}

One can estimate all the above mentioned substitution frequencies from real
sequence data by comparing a pair of ancestral
$\vec{\alpha}=\alpha_1\alpha_2\dots\alpha_N$ and daughter sequence
$\vec{\beta}=\beta_1\beta_2\dots\beta_N$, where the daughter sequence
represents the state of the ancestral sequence after the substitution processes
acted upon it for some time. Note that we do not assume any other properties
regarding to the nucleotide or dinucleotide distributions of the sequences.
Especially, the two sequences do not need to be in their stationary state with
respect to the substitution model. [In practice, these pairs of ancestral and daughter
sequences can be obtained in various ways. One very fruitful approach is to
take alignments of repetitive sequences, which can be found in various genomes
due to the activity of retroviruses.  Such repetitive elements have entered
these genomes during short periods in evolution.  Hence all copies of such
elements in a genome have been subject to nucleotide substitutions for the same
time and accumulated corresponding amounts of changes.  Various such repetitive
elements and their respective alignment to the once active master (which is
taken to be the ancestral sequence \citep{APH03}) can be identified using the
RepeatMasker, http://www.repeatmasker.org.]

The log likelihood that a sequence
$\vec{\beta}$ evolved from a master sequence $\vec{\alpha}$ under a given
substitution model parameterized by the substitution frequencies $\{r\}$ is
given by 
\begin{eqnarray} 
\log L_{\{r\}}&=&
\log P_{\{r\}}(\vec{\beta}|\vec{\alpha}) 
\nonumber\\
&\approx&
\log \prod_{i=2}^{L-1}
P_{\{r\}}(\cdot\beta_i\cdot|\alpha_{i-1}\alpha_i\alpha_{i+1}) 
\nonumber\\
&=&
\sum_{\alpha_1\alpha_2\alpha_3\beta_2}
N(\alpha_1\alpha_2\alpha_3\ra\cdot\beta_2\cdot)
\log
P_{\{r\}}(\cdot\beta_2\cdot|\alpha_1\alpha_2\alpha_3)
\;.
\label{eqll}
\end{eqnarray} 
where  $P_{\{r\}}(\vec{\beta}|\vec{\alpha})$ is the probability of the
evolution of the sequence $\vec{\alpha}$ into $\vec{\beta}$.  This probability
can very well be approximated by the product in the second line. 
This is due to the fact that the correlations induced by the substitutional 
processes are very short ranged \citep{ABH02}. We therefore 
take into account the identities of bases and the dynamics on
the nearest neighbors to the left and to the right, and neglect
those on the next nearest neighbors and beyond.
For most applications
this approximation turns out to be sufficient since estimated 
substitution frequencies deviate less than 1\% from their actual
values (see below).
Note that this approximation is even exact in the absence of neighbor dependent 
substitution processes. The numbers 
$N(\alpha_1\alpha_2\alpha_3\ra\cdot\beta_2\cdot)$ denotes the 
counts of observations of a base substitution from $\alpha_2$ 
(flanked by $\alpha_1$ to the
left and $\alpha_3$ to the right) to $\beta_2$. 

To estimate the substitution frequencies $\{r^\star\}$
for a given pair of $\vec{\alpha}$ and
$\vec{\beta}$ or given numbers $N(\alpha_1\alpha_2\alpha_3\ra\cdot\beta_2\cdot)$ we 
have to maximize the above likelihood by adjusting the
substitution frequencies.  This can easily be done using Powell's method
\citep{Pr92} while taking care of boundary conditions \citep{Bo66}, i.e. the
positivity of the substitution frequencies. 

\begin{figure}[htb]
\bigskip
\bigskip
  \begin{center}
    \includegraphics[width=0.8\textwidth]{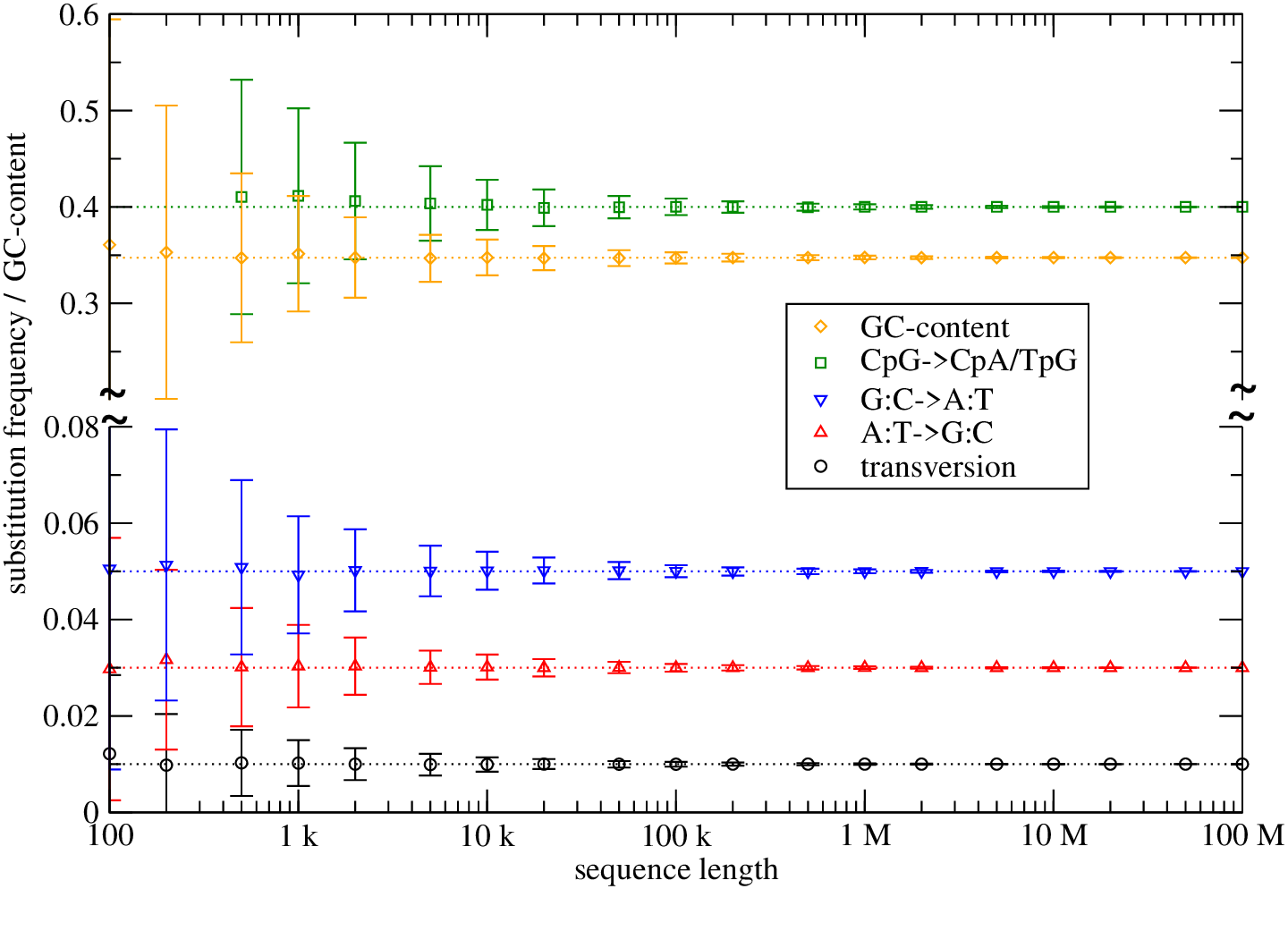}
	\caption{\footnotesize
\label{fig1}Plot of the estimated frequencies and their standard
deviation (from 500 measurements) for randomly drawn sequences of various
length.  The daughter sequences have been synthetically aged using the
following processes (with frequency as indicated by the dotted lines):
transversions (0.01), \A:\T\rat\G:\C\ (0.03), \G:\C\rat\A:\T\ (0.05), and
\CpG\rat\CpA/\TpG\ (0.4). The stationary GC-content for this model is $0.3474$.
}
  \end{center}
\end{figure}

\begin{figure}[htb]
\bigskip
\bigskip
  \begin{center}
    \includegraphics[width=0.8\textwidth]{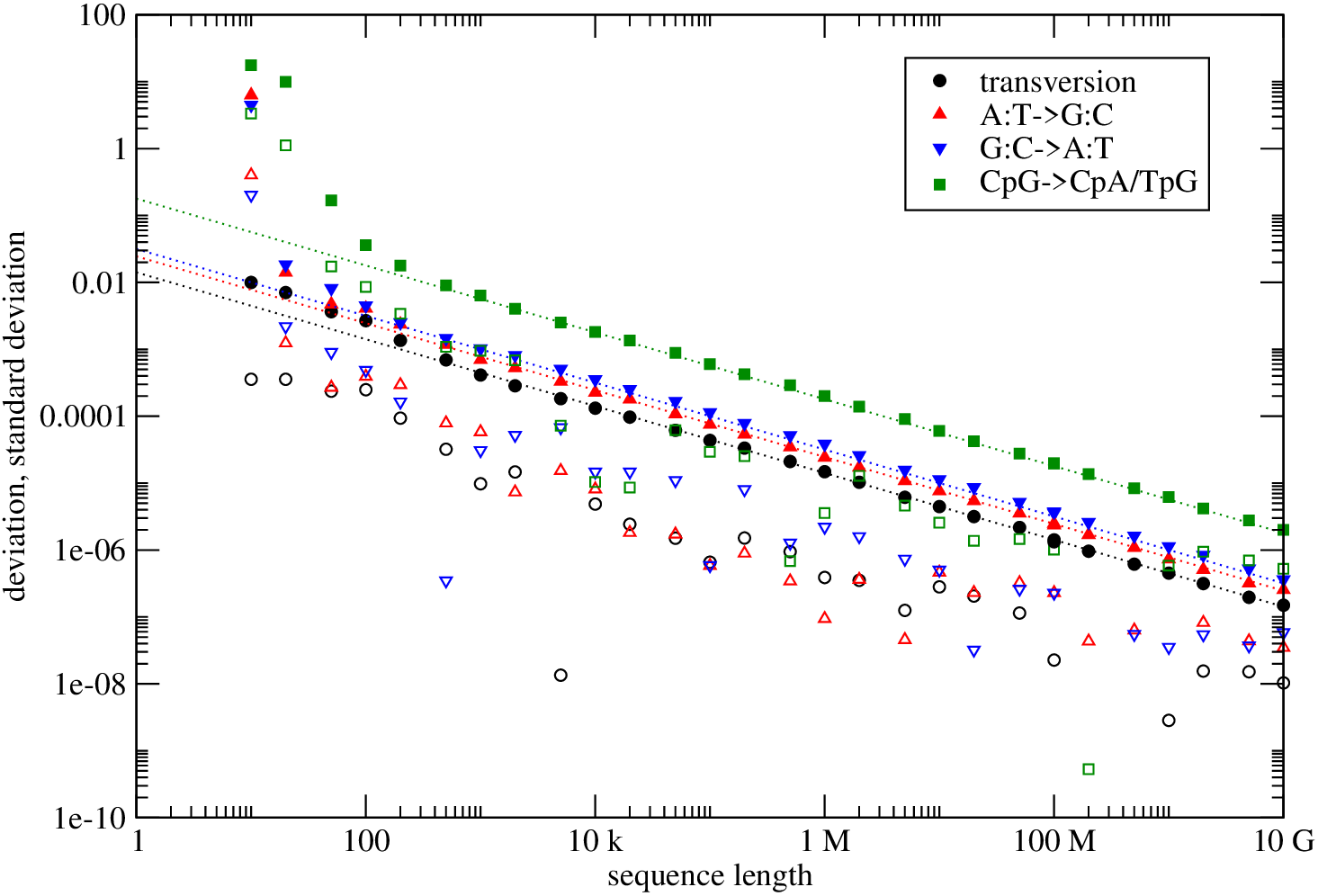}
	\caption{\footnotesize\label{fig2}Plot of the deviations of the estimated frequencies
$\{|\bar{r}^*-\hat{r}|\}$ (open symbols) and the standard deviation $\{\Delta
r^*\}$ (closed symbols) from 500 measurements for randomly drawn sequences of
various lengths.  The daughter sequences have been synthetically aged using the
following processes (with frequency): transversions (0.0001), \A:\T\rat\G:\C\
(0.0003), \G:\C\rat\A:\T\ (0.0005), and \CpG\rat\CpA/\TpG\
(0.004).
}
  \end{center}
\end{figure}

\begin{figure}[htb]
\bigskip
\bigskip
  \begin{center}
    \includegraphics[width=0.8\textwidth]{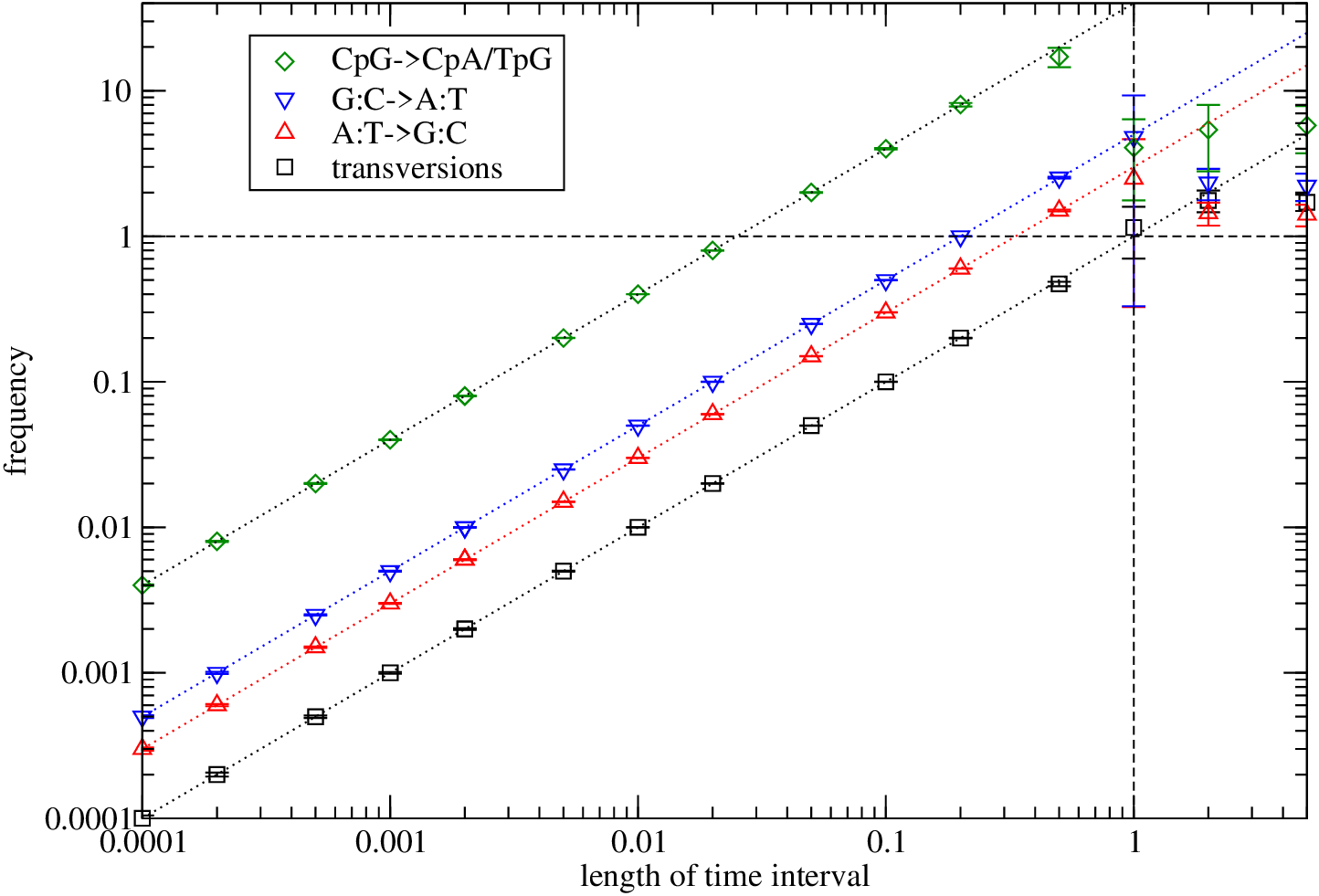}
	\caption{\footnotesize\label{fig3}A plot of the estimated frequencies for various
degrees of sequence divergence. The dotted lines give expected values of the
frequencies. The sequence length has been chosen to be $N=10^7$. }
  \end{center}
\end{figure}

\subsection{Uncertainty of estimates for finite sequence length}

Due to the stochastic nature of the substitution process and due to the fact
that always only a finite amount of sequence data is available to estimate
the substitution frequencies $\{r^\star\}$, estimated frequencies will show
deviations from the real substitution frequencies.  In general we do not know
or cannot infer these real frequencies otherwise.  In order to be able to
analyze the uncertainty of frequency estimates from finite sequences we
synthetically (in silico) generate pairs of ancestral and daughter sequences
using known substitution processes and rates $\{\hat{r}\}$.  In the following
section we include just one neighbor dependent substitution process, namely the
\CpG-methylation deamination process, \CpG\rat\CpA/\TpG, which plays a
predominant role in the analysis of nucleotide substitutions in vertebrates.
The nucleotides of the ancestral sequences $\vec{\alpha}$ (of length $N$) have
been chosen randomly with equal probability from the 4 nucleotides.
Subsequently, the ancestral sequence was synthetically aged and we applied
substitutions using a Monte Carlo algorithm as described in \citep{ABH02}
yielding the sequence $\vec{\beta}$.  The resulting pair of sequences is then
analyzed using the above procedure to get estimates of the rates $\{r^\star\}$.
We repeated this experiment 500 times and got estimates for the means
$\{\bar{r}^*\}$ and standard deviation $\{\Delta r^*\}$ of these measurements.
In addition we computed the stationary GC-content from each set of substitution
frequencies \citep{ABH02}. Results of this analysis are presented in
Figure~\ref{fig1} where we show the mean and standard deviation of estimated
rates for different length of sequences $N$.  The transversion frequencies were
chosen to be 0.01, the frequency of the \A:\T\rat\G:\C\ transition to be 0.03,
that of the \G:\C\rat\A:\T\ transition to be 0.05, and that of the
\CpG\rat\CpA/\TpG\ transition to be 0.4, as indicated by the doted lines in
Figure~\ref{fig1}. This choice of frequencies mimics the relative strength of
the substitution process as they are observed in the human genome.  As can be
seen the uncertainty of observed substitution frequencies correlates positively
with the substitution frequencies and negatively with the length of the
sequences.

To further quantify these uncertainties and discuss their dependence on various
quantities we plotted the deviations $\{|\bar{r}^*-\hat{r}|\}$ and the standard
deviations $\{\Delta r^*\}$ as a function of the sequence length $N$ in
Figure~\ref{fig2}.  The standard deviations decrease with $1/\sqrt{N}$. In the
absence of neighbor dependent substitutions and for ancestral sequences with
equally probable nucleotides the standard deviation for reverse complement
symmetric frequencies can actually be calculated to be
\begin{equation}
\Delta r^*_{\alpha\beta}=
\left(\frac{2 r_{\alpha\beta}}{N}\right)^{1/2}
\label{dri}
\end{equation}
as long as all frequencies $r\ll  1$.
Corresponding lines are presented also in Figure~\ref{fig2} and fit the observed
deviations well. The deviation for neighbor dependent processes 
such as the process \CpG\rat\CpA/\TpG\ can be computed to be of
the order of:
\begin{equation}
\Delta r^*_{\alpha\beta\gamma\delta}=
\left(\frac{8 r_{\alpha\beta\gamma\delta}}{N}\right)^{1/2}
\label{drii}
\end{equation}
Note, that for $r\ll 1$ these errors stem only from the stochastic nature of
the underlying substitutional process and are not due to approximations used
during our maximum likelihood analysis of the sequence pairs $\vec{\alpha}$ and
$\vec{\beta}$ as described in the previous section.

The deviations of the observed from the real frequencies
$\{|\bar{r}^*-\hat{r}|\}$ (see Figure~\ref{fig2}) also decrease with
$1/\sqrt{N}$ and are always bounded from above by  $\{\Delta r^*\}$.  Note,
that the estimates of substitution frequencies are very precise, although we
used an approximation when deriving the likelihood in Eq. (\ref{eqll}).  This
property does not hold true for neighbor dependent processes in general. For
instance, we observe small (below 1\%, data not shown) but systematic
deviations of the estimated substitution frequencies if we include the process
\ApA/\TpT\rat\CpA/\TpG. In this case, one should also take into account the
identity and dynamics of nucleotides on next nearest neighbor sites and the
associated neighbor dependent processes.  One would have to introduce higher
order corrections in Eq. (\ref{eqll}).  This is true because of overlapping
initial states of the  neighbor dependent process, i.e.  two \ApA's in a
triplet \A\A\A.  However, such corrections do not have to be considered for the
\CpG\rat\CpA/\TpG\ process.  For a given \CpG, the next nearest neighbor
dependent process might only occur on a neighboring \CpG, which in contrast to
\ApA's cannot overlap with the given \CpG.  Hence correlations to the next
\CpG\ are even smaller, which makes the estimation of substitution frequencies
neglecting such correlations very precise.  In the absence of any neighbor
dependent process there is no approximation involved to compute the likelihood
in Eq. (\ref{eqll}) and therefore estimates will be asymptotically exact for
$N\ra\infty$.

The above formulas for the standard deviation, Eqs. (\ref{dri}) and
(\ref{drii}), lose their validity if any one of the frequencies is of the
order of one. However, the standard deviations are still decreasing with
increasing sequence length. In Figure~\ref{fig3} we present estimated
frequencies from sequences of various degrees of divergence.  The substitution
rates have been chosen in the ratios 1:3:5:40 for the transversions, the
\A:\T\rat\G:\C\ transition, the \G:\C\rat\A:\T\ transition, and the
\CpG\rat\CpA/\TpG\ process. On the horizontal axis we plot the length of the
time interval the ancestral sequenced (of length $N=10^7$) has been aged. The
dotted lines give the real substitution frequencies, which are the products of
the corresponding rates and the length of the time interval.  As long as not
all substitution frequencies are greater than one (to the left of the dashed
vertical line in Figure~\ref{fig3}) the substitution frequencies can
faithfully estimated, even if single frequencies exceed one (the dashed
horizontal line).  If all substitution frequencies are of the order of or
larger than one, the estimation of substitution frequencies is not possible
anymore (to the right of the dashed vertical line).  In this case, more or less
all nucleotides underwent one or more substitution processes making it
impossible to estimate the frequencies of the underlying processes.

In reality however, the nucleotides in the ancestral sequence will not be
randomly distributed with equal probability from the 4 nucleotides (as assumed
above). On top of that genomic sequences will show non-trivial dinucleotide
distributions, i.e. neighboring bases are not independent and the dinucleotide
frequencies $f_{\alpha\beta}$ will deviate from the product of nucleotide
frequencies $f_\alpha f_\beta$ \citep{CB95}.  Both these factors will influence
the deviations between the observed and the real substitution frequencies and
in those cases the above formulas (\ref{dri}) and (\ref{drii}) do not hold
anymore.  We also expect additional errors due to the presence of unaccounted
neighbor dependent processes.  Depending on the magnitude of the rates for such
processes the errors can get quite significant as discussed below. To exclude
the latter type of errors one actually has to try to incorporate additional
neighbor dependent processes and judge whether their inclusion is actually
relevant (as discussed in the next subsection).

For genomic applications, it is further not possible to repeat the measurements
of substitution frequencies for different sets of sequences to get an estimate
of the typical errors.  However, one can still get estimates on the expected
standard deviation from bootstrapping the available data. One has to resample
the available data drawing randomly and with replacement $N$ pairs of aligned
ancestral and daughter nucleotides (keeping the information of the ancestral
base identity to the left and to the right) and generate a list of counts
$N(\alpha_1\alpha_2\alpha_3\ra\cdot\beta_2\cdot)$ which then will be used to
maximize the likelihood and estimate the substitution frequencies as described
above.  One  repeats this resampling procedure $M$ times and from the $M$
estimates of the substitution frequencies and stationary GC-content calculates
their standard deviation, which gives the statistical error due to the limited
amount of sequence data. We found that $M = 500$ samples are sufficient to
estimate those errors (data not shown).

\subsection{Extending the model to include additional processes}

Next we address how one can extend 
a given substitution model and
include additional neighbor dependent processes to maximize the potential of
such a model to describe the observed data. 
With the inclusion of additional neighbor dependent processes the likelihood of
a model $\{r'\}$ will in any case be greater than the one of the original model $\{r\}$.
This is true because the models are nested and one has one more free parameter 
to explain the given data. 
To test whether the inclusion of a new parameter is justified we employ
the likelihood ratio test for nested models. Let
$\lambda=L_{\{r\}}/L_{\{r'\}}$ be the likelihood ratio, then $-2\log\lambda$ has
an asymptotic chi-square distribution with degrees of freedom equal to the
difference in the numbers of free parameters of the two models, which in our
case is one \citep{EG01}.

In practice we extend a given substitution model in turn by one out of the
$4\times 4\times 3\times 2=96$ possible neighbor dependent processes.  Out of
those extended models we choose the best one, i.e. the one with the highest
likelihood $L_{\{r'\}}$.  Since the best is chosen out of a finite set of possibilities,
we have to account for multiple testing and use a Bonferroni
correction.  Hence we require that $-2\log\lambda>15$ to have significance on
the 5\% level\footnote{Note that $\int_0^{15}
\chi^2_1(x)\,dx=0.99989>1-0.05/96$}.  We confirmed this conservative threshold
also by simulations using sequences that have been synthetically mutated
according to a known model.

\section{Results}
\tabi
As a first test, we applied the described method to 
identify and measure neighbor dependent substitution processes 
to human genomic data. We
took the copies of the AluSx SINEs that have been found in a genome-wide search
of the human genome (release v20.34c.1 at ensembl.org from April 1st, 2004).
These elements are assumed to have evolved neutrally and therefore the
substitution process is reverse complement symmetric. Results are presented in
Table 1.
In the first column of data we give estimations for the 6 neighbor independent
single nucleotide substitutions. We subsequently tested 48 possible
extension of this simple substitution model by one additional neighbor
dependent substitution process together with its reverse complement symmetric process
(Note that in this case only 48 extensions have to be considered).
As
expected (and shown in the second column in Table 1) the \CpG\ methylation
deamination process (\CpG\rat\CpA/\TpG) turns out give the best improvement
with $-2\log\lambda=7.7\cdot 10^6$, which is clearly above the threshold of
$15$.  The substitution frequency of this process is about 45 times higher than
that of a transversion.  Extending the model from 6 to 7 parameters and
including the \CpG\rat\CpA/\TpG\ process, mostly affects the estimate for the
\G:\C\rat\A:\T\ transition, which decreases about a factor three.  Please also
note that subsequently the estimation of the stationary GC-content from those
rates rises from 21\% for the 6 parameter model to 34\% for the 7 parameter
model. This reveals that estimates of 
substitution frequencies and 
 the stationary nucleotide composition are
very much affected by the underlying substitution model.
Substantial deviations can be observed when 
the substitution model does not include all relevant process, as it the case
for the 6 parameter model for nucleotide substitutions in the human lineage.
In principle there can be even more neighbor dependent processes, which we have to
account for. We therefore try to incorporate an additional process
besides the already found one.

The second process that needs to be included to improve the model is the
substitution of \CpG\rat\CpC/\GpG\ ($-2 \log\lambda=1.3\cdot 10^5$). This is
another \CpG\ based process and probably also triggered by the methylation of
cytosine. However, the substitution frequency is about 30 times smaller than
this of the \CpG\rat\CpA/\TpG\ process. The third process is then the
substitution \TpT/\ApA\rat\TpG/\CpA\ ($-2\log\lambda=9.6\cdot 10^4$). The
instability of the \TpT\ dinucleotide does not come as a surprise here, since
two consecutive thymine nucleotides tend to form a thymine photodimer
$\T\!<>\!\T$. This process is one of the major lesions formed in DNA during
exposure to UV light \citep{DZC97}.

\smallskip

Next we turn to the analysis of the DANA repeats in zebrafish ({\em Danio rerio}).
Results are presented in Table 2. Again we start with a model just comprising
single nucleotide transversions and transitions. As observed in human the
transitions occur more often than transversions and there is a strong \A:\T\ bias
in the single nucleotide substitutions. Zebrafish being a vertebrate also
utilizes methylation as an additional process to regulate gene expression. As a
consequence we observe a higher mutability of the \CpG\ dinucleotide due to the
deamination process also in zebrafish. However the substitution frequency for
the \CpG\rat\CpA/\TpG\ process is in zebrafish only about 8 times higher than this of
a transversion suggesting that the degree of methylation is generally lower
than in human. 

\tabii
\smallskip

We also investigated non-vertebrate sequence data. As an example we
present here the analysis of the DNAREP1\_DM repeat in {\em Drosophila melanogaster}
(Table 3). The case to include neighbor dependent process is in this clearly
not as strong as for vertebrate genomes. The values of $-2\log\lambda$ are 3 orders of
magnitude smaller but still above threshold for the first 3 processes which are
chosen by our procedure to be included into a model for nucleotide
substitutions in fly. The first such process is the substitution \TpA\rat\TpT/\ApA.
Although the corresponding substitution frequency is lower than all the single
nucleotide transitions and transversions, the dinucleotide frequencies in the
stationary state deviate up to 10\% from their neutral expectation under a
neighbor independent substitution model (data not shown). Therefore even processes with
a small contribution to the overall substitutions have a large influence on the
observed patterns of dinucleotide frequencies or genomic signatures and
therefore may very well be solely responsible for the generation of such
pattern in different species.

\tabiii

\section{Conclusion}

We presented a framework to identify the existence and measure the rates of
neighbor dependent nucleotide substitution processes.  We discussed the
extension of models of nucleotide substitutions in human and included more
neighbor dependent processes besides the well-known \CpG\ methylation
deamination process \citep{ABH02}. We could also show that the \CpG\
methylation deamination is the predominant substitution process in zebrafish,
while it does not play a role in fruit fly. We exemplified our method 
using sequence data from one particular subfamily of repeats from these three
organisms. In the case of the human genome a much more thorough analysis on
various families of repeats have been presented in \citep{APH03}.  A similar
study, which also would have to include also neighbor dependent substitutions, for
other species will further broaden our knowledge about the molecular processes
that are responsible for nucleotide mutations and their fixation. 

{\bf Acknowledgment}
We thank Nadia Singh and Dmitri Petrov (Stanford) for kindly
providing sequence data on the DNAREP1\_DM repeat in {\em Drosophila
melanogaster}.

\newpage
\def\etal{{\em et.al.}}


\begin{thebibliography}{}

\bibitem[Arndt \etal, 2002]{ABH02}
Arndt, P. F., Burge, C. B. and Hwa, T. (2002). 
DNA Sequence Evolution with Neighbor-Dependent Mutation.
6th Annual International Conference on Computational Biology RECOMB2002, Washington DC, ACM Press, KK.

\bibitem[Arndt \etal, 2003]{APH03}
Arndt, P. F., Petrov, D. A. and Hwa, T. (2003). 
Distinct changes of genomic biases in nucleotide substitution at the time of Mammalian radiation.
{\em Mol Biol Evol} {\bf 20}(11): 1887-96.

\bibitem[Box, 1966]{Bo66}
Box, M. J. (1966). 
A Comparison of Several Current Optimization Methods and Use of Transformations in Constrained Problems.
{\em Computer Journal} {\bf 9}(1): 67-77.

\bibitem[Coulondre \etal, 1978]{Co78}
Coulondre, C., Miller, J. H., Farabaugh, P. J., et al. (1978). 
Molecular basis of base substitution hotspots in Escherichia coli.
{\em Nature} {\bf 274}(5673): 775-80.

\bibitem[Douki \etal, 1997]{DZC97}
Douki, T., Zalizniak, T. and Cadet, J. (1997). 
Far-UV-induced dimeric photoproducts in short oligonucleotides: sequence effects.
{\em Photochem Photobiol} {\bf 66}(2): 171-9.

\bibitem[Ewens and Grant, 2001]{EG01}
Ewens, W. J. and Grant, G. (2001). 
{\em Statistical methods in bioinformatics : an introduction.}
New York, Springer.

\bibitem[Karlin and Burge, 1995]{CB95}
Karlin, S. and Burge, C. (1995). 
Dinucleotide relative abundance extremes: a genomic signature.
{\em Trends Genet} {\bf 11}(7): 283-90.

\bibitem[Karlin and Mr\'azek, 1997]{KM97}
Karlin, S. and Mr\'azek, J. (1997). 
Compositional differences within and between eukaryotic genomes.
{\em Proc Natl Acad Sci U S A} {\bf 94}(19): 10227-32.

\bibitem[Karlin \etal, 1997]{KMC97}
Karlin, S., Mr\'azek, J. and Campbell, A. M. (1997). 
Compositional biases of bacterial genomes and evolutionary implications.
{\em J Bacteriol} {\bf 179}(12): 3899-913.

\bibitem[Lio and Goldman, 1998]{LioGoldman}
Lio,P. and Goldman,N. (1998). 
Models of molecular evolution and phylogeny. 
{\em Genome Res.}, {\bf 8}, 1233-1244.

\bibitem[Lunter and Hein, 2004]{LH04}
Lunter, G. and Hein, J. (2004). 
A nucleotide substitution model with nearest-neighbour interactions.
{\em Bioinformatics} {\bf 20} Suppl 1:I216-I223.

\bibitem[Press \etal, 1992]{Pr92}
Press, W. H., Teukolsky, S. A., Vetterling, W. T., et al. (1992).
{\em Numerical Recipes in C, The art of scientific computing.}
Cambridge, Cambridge University Press.

\bibitem[Razin and Riggs, 1980]{RR80}
Razin, A. and Riggs, A. D. (1980). 
DNA methylation and gene function.
{\em Science} {\bf 210}(4470): 604-10.

\bibitem[Russell \etal, 1976]{Ru76}
Russell, G. J., Walker, P. M., Elton, R. A., et al. (1976). 
Doublet frequency analysis of fractionated vertebrate nuclear DNA.
{\em J Mol Biol} {\bf 108}(1): 1-23.

\bibitem[Russell and Subak-Sharpe, 1977]{RS77}
Russell, G. J. and Subak-Sharpe, J. H. (1977). 
Similarity of the general designs of protochordates and invertebrates.
{\em Nature} {\bf 266}(5602): 533-6.

\bibitem[Siepel and Haussler, 2004]{Haussler}
Siepel, A. and Haussler, D. (2004).
Phylogenetic estimation of context-dependent substitution rates by maximum likelihood.
{\em Mol Biol Evol.} {\bf 21}(3):468-88.




\end{thebibliography}
\end{document}